\documentclass[referee]{aa}
\usepackage{amsmath}
\usepackage{graphicx}
\usepackage{txfonts}

\begin{document}

   \title{Origin of  craters on Phoebe: comparison with Cassini's data.}

   \author{R. P. Di Sisto 
           \thanks{romina@fcaglp.unlp.edu.ar}
           and A. Brunini
      }

   \offprints{R. P. Di Sisto
    }

  \institute{Facultad de Ciencias Astron\'omicas y Geof\'\i sicas, Universidad 
Nacional de La Plata and \\
Instituto de Astrof\'{\i}sica de La Plata, CCT La Plata-CONICET-UNLP \\
   Paseo del Bosque S/N (1900), La Plata, Argentina.
                }

   \date{Received / Accepted}


\abstract
{ Phoebe is one of the irregular satellites of Saturn; the images taken by Cassini-Huygens 
spacecraft allowed us to analyze its surface and the craters on it.}
{We study the  craters on Phoebe produced by Centaur objects from the Scattered Disk (SD)
and plutinos escaped from the  3:2 mean motion resonance with Neptune and compare 
our results with the observations by Cassini.}
{ We use previous simulations on trans-Neptunian  Objects and a method that 
allows us to obtain the number of craters and  the cratering rate on Phoebe.}
{We obtain  the number of craters and the greatest crater on Phoebe
produced by Centaurs in the present configuration of the Solar System. 
Moreover, we obtain a present 
normalized rate of encounters of Centaurs with Saturn of $\dot F = 7.1 \times 10^{-11}$
per year, from which we can infer the current cratering rate on Phoebe for each crater diameter.}
{Our study and the  comparison with the observations suggest that the main 
crater features on Phoebe are unlikely to have been produced  in the present 
configuration of the Solar System and that they must have been   
acquired when the SD were depleted in the early  Solar System. 
If this is what happened and the craters were produced when Phoebe was a satellite of Saturn, 
then it  had to be captured, very early in the  evolution of the Solar System.} 


\authorrunning{R. P. Di Sisto \& A. Brunini 
                              }
\titlerunning{Origin of the craters on Phoebe }

\maketitle

\section{Introduction}

Phoebe is one of the irregular satellites of Saturn. It has a retrograde orbit, which suggests 
that it was captured by Saturn instead of being formed ``in situ'' (e.g Pollack et al.
\cite{Pollack79}).  Moreover, Phoebe's composition is close to that derived from bodies such 
as Triton and Pluto, and it is different from that of the regular satellites of Saturn 
supporting Phoebe's origin as a captured body of the outer Solar System 
(Johnson \& Lunine \cite{JL05}). 

On 11 June 2004, Cassini-Huygens spacecraft  encountered Phoebe in a fly-by within 2000 km 
of Phoebe's surface. This encounter allowed  Cassini to analyze Phoebe's surface in detail 
which improved  the previous Voyager data.   
 Buratti et al. (\cite{Buratti08}) analyzed  and characterized the physical properties 
of the surface from photometric data from Cassini VIMS (Visual and Infrared Spectrometer), concluding 
that it is rough and dusty, perhaps from a history of out-gassing or a violent collisional 
history suggested by Nesvorny et al. (\cite{Nesvorny03}).  
Hendrix \& Hansen  (\cite{HH08}) analyzed the first UV spectra of Phoebe with the 
Cassini Ultraviolet Imaging Spectrograph (UVIS) during the Cassini spacecraft fly-by 
and  detected water ice. Using VIMS data, Cruikshank et al. (\cite{C08}) concluded that Phoebe's 
surface is rich in organics, which is compatible with the low albedo of Phoebe.   
Porco et al (\cite{Porco05}) calculated a mean density of Phoebe of $1.63 gr/cm^3$ from calculation  
of the volume and the determination of the mass from tracking the spacecraft. If   
Phoebe's surface was a mixture of rock and ice, the previous density  is 
compatible with a porosity lower than $\sim 40 \%$  (Porco et al.  \cite{Porco05}).
Johnson and Lunine (\cite{JL05}) analyzed the relation between composition and probable 
porosity of Phoebe and they found that if Phoebe was derived from the same compositional 
reservoir as Pluto and Triton, Phoebe's measure density is consistent with a porosity of 
$\sim 15 \%$.

Giese et al. (\cite{Giese06}) presented the results of a photogrametric analysis of the 
high-resolution stereo images of Phoebe. In particular they obtained a mean figure radius 
of $107.2$ km and  a digital terrain model of the surface reveals significant morphological 
detail. 
The images revealed that Phoebe basically exhibits simple crater shapes with the only 
exception of the greatest impact crater Jason with a diameter of $\sim 100$ km. 
Several of the smaller craters present pronounced conical shapes which could indicate the presence 
of porous, low compacting material on the surface of Phoebe.

Kirchoff \& Schenk (\cite{KS10}) reexamined the impact of crater distribution of the mid-sized 
saturnian satellites. For Phoebe they found that the crater size frequency distribution has 
relatively constant values for crater diameters $D \leq 1$ km, but then it has a sudden and confined 
dip around $D \sim 1.5$ km. Beyond this dip, the crater size frequency distribution has a slow 
increase. This behaviour is unique in the saturnian satellite system and it is probably connected 
with Phoebe's origin.   

Zhanle et al. (\cite{Z03}) write a previous paper that calculates cratering rates in the satellites 
of the outer planets. They used impact rates on the giant planets obtained by Levison 
\& Duncan (\cite{LD97}) and independent constraints on ecliptic comets. Their results will be 
compared with ours.  

As we have seen, the origin of craters on Phoebe is unclear. 
But the main population of objects that can produce craters on Phoebe are Centaurs, since 
they are the small body objects that cross the orbit of the giant planets, in particular 
the orbit of Saturn, and then its satellites.

Centaurs are transient bodies between their source in the trans-Neptunian population and the Jupiter 
Family Comets. They come mainly from a sub population in the trans-Neptunian zone, the Scattered Disk 
Objects (SDOs). The SDOs are bodies with perihelion distances $q$ greater than 30 AU and smaller than 
$\sim 39$ AU that can cross the orbit of Neptune and eventually evolve into the giant planetary zone, 
crossing the orbits of those planets, and then the orbits of their satellites (Di Sisto \& Brunini 
\cite{Disisto07},  Levison \& Duncan \cite{LD97}). 
The secondary source of Centaurs are  plutinos and the low eccentricity trans-Neptunian objects 
(Di sisto et al. \cite{Disisto10}, Levison \& Duncan \cite{LD97}). 
Plutinos are those trans-Neptunian objects located in the 3:2 mean 
motion resonance with Neptune at $a \sim 39.5$ AU. They are ``protected'' by the 
  3:2 mean motion resonance with Neptune  
but some of them are long term escapers that are presently escaping from the resonance 
(Morbidelli \cite{Morby97}).  In this paper we will
study the production of craters on Phoebe from Centaur objects from SDOs and plutinos escaped 
from the  3:2 mean motion resonance with Neptune, as the two main populations of impactors. 
We use here previous 
simulations on trans-Neptunian  Objects (TNOs) and a method that allows us to obtain 
directly the cratering rate on Phoebe.
This study, and the comparison with the observations of Cassini images  may help us 
to determine the origin of crates on Phoebe, in order to determine the history of Phoebe's surface 
and also constrain its origin.

\section{The number of SDOs}
\label{ssdo}

Cratering rates depend on the number and sizes of the impactor population. Thus 
 we must know the real initial number of SDOs to calculate the total number of 
collisions on Phoebe.
Then, we are going to estimate the total number of present SDOs.

f Parker \& Kavelaars (\cite{PK10a}) re-characterized the orbital sensitivity of 
several published pencil-beam surveys. They found that these surveys were sensitive to 
 distant populations like SDOs and  Sedna-like objects. Using this result Parker \& Kavelaars 
(\cite{PK10b}) derived new upper limits on those distant populations 
and  used this new limits to obtain the number of SDOs. To do this 
they performed a model that considered two laws for the radial distance distribution of SDOs. On 
the one hand they took a radial distance distribution of SDOs $\propto r^{-1.5}$ and obtained 
a maximum population of $ N(d > 100$ km $) = 3.5 \times 10^5 $. On the other hand they took 
a uniform radial distance distribution obtaining in this case a maximum population 
of $ N(d > 100$ km $) = 25 \times 10^5 $.
 In this paper we  take the number of SDOs greater than $d = 100$ km equal to
 $ N(d > 100$ km $) = 3.5 \times 10^5 $ since the considered radial distance distribution  
is consistent with the one obtained by Di Sisto \& Brunini  (\cite{Disisto07}). 
Then  the total population of SDOs with diameter greater than $d_0$ will 
be given by $N(d > d_0) = 3.5 \, \times \,10^5 \,\,(d_0/100)^{-s+1}$, where 
 $d_0$ must be in km and $s$ is the index of the differential size distribution. 
There are some authors who have found a single power law size distribution 
for TNOs (Petit et al. \cite{Petit06}, Fraser et al. \cite{F08}). 
However, other papers  suggest that the size distribution function  (SDF) of TNOs could 
have a break at a diameter  of $\sim 60$ km  
(Bernstein et al. \cite{Bernstein04},  Gil Hutton et al. \cite{GH09},  Fraser \& Kavelaars 
\cite{FK09}, Fuentes \& Holman \cite{FH08}, Fuentes et al. \cite{F09}).
The differential power law indexes for smaller TNOs (this is $d < 60$ km) found by those surveys 
are $s_2 = 2.8, 2.4, 1.9, 2.5$ and $2$, respectively. 
It seems to be enough evidence for a break in the {}size distribution of the TNO 
population. In particular we are going to assume that this break is also valid for all the 
dynamical classes of TNOs. 

 Elliot et al.  (2005) accounts for the  SDF for each dynamical 
 class in the TN region. 
Specifically for SDOs they found that the differential size distribution index is $s_1 = 4.7$ 
 for the  brightest objects.
Then taking into account the assumed break in the  SDF of SDOs
we consider that the  power law  SDF of SDOs  breaks at $d \sim 60$ km to an index 
 of between $3.5 $ and $ 2.5$. 
We analyze those indexes as limit cases that give higher and lower ranges for the population 
of SDOs and then the production of craters on Phoebe. 
The higher value of $s_2 = 3.5$ corresponds to a population in steady-state 
(Dohnanyi 1969) which could be the case for the smallest SDOs 
(Gil Hutton et al. \cite{GH09}). 
Considering all this, the number of SDOs greater than a  diameter $d_0$ will be 
given by 
\begin{xalignat}{4}
N(d>d_0) &= C_0 \bigg(\frac{1 \text{km}}{d}\bigg)^{s_2 - 1} &&\text{for} && d \leq 60 \text{km},  \nonumber \\
N(d>d_0) &= 3.5 \times 10^{5} \bigg(\frac{100 \text{km}}{d}\bigg)^{s_1-1} &&\text{for} && d > 60 \text{km},
\label{nr}
\end{xalignat} 
where $C_0 = 3.5 \times 10^{5} 100^{s_1-1} (60)^{s_2-s_1}$ by continuity for $d = 60$ km, 
$s_1 = 4.7$ and $s_2 = 2.5$ and $3.5$.
This law is plotted in Fig. 1 with the two breaks considered. 
\begin{figure}
\centering
\resizebox{\hsize}{!}{\includegraphics{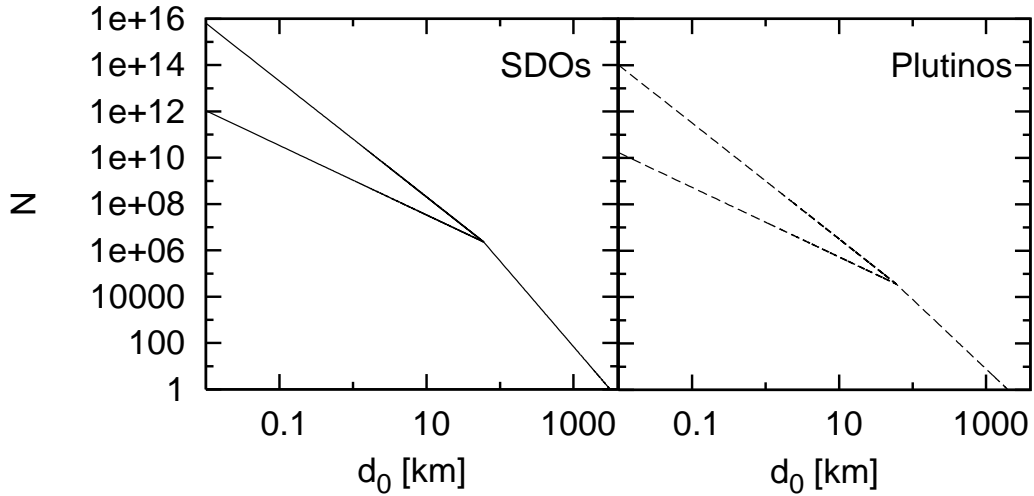}}
\caption{ Cumulative number of SDOs and Plutinos according 
to the size distribution laws described in the text. }
\label{fig1}
\end{figure}

\section{SDO collisions on Phoebe}
\label{sdo-col}

To study the collisions of SDOs on Phoebe and the contribution of that population 
to the cratering history of the satellite we use some of the outputs of the 
numerical simulation performed in a previous paper by 
Di Sisto \& Brunini (\cite{Disisto07}). In that work we integrated  numerically 
1000 objects from the SD 
(95 real + 905 fictitious) and studied their evolution in the Centaur zone 
under the gravitational action of the Sun and  the four giant planets. The 
computations were followed for $4.5$ Gyr, or until the test body collided 
with a planet, was ejected, or entered the region inside Jupiter's orbit 
($r<5.2$ AU). 
In that paper we also stored in a file the encounters of the fictitious SDOs with the 
planets and registered the time of the encounter, the minimum distance to the planet ($q$)
and the relative velocity at this distance ($v(q)$). From these data 
we can calculate, the number of encounters with Saturn within the Hill's sphere of
the planet. 
Using the particle in a box approximation and assuming that the geometry of the encounters is 
isotropic it is possible to calculate the number of collisions on Phoebe ($N_c$) 
 through the relation: 
\begin{equation}
\frac{N_{c}}{N_e} = \frac{v_i \, \,R_p^2}{v(R)\,\, R^2}
\label{nc}
\end{equation}
Where $N_e$ is the number of encounters with Saturn inside its Hill's sphere of radius 
$R$, $R_p$ is the radius of Phoebe, $v(R)$ is the mean relative encounter velocity of SDOs 
when entering the Hill's sphere of the planet and $v_i$ is the collision velocity of 
SDOs on Phoebe. 
$v(R)$ can be calculated from $v(q)$ registered in our outputs from 
the relation: 
\begin{equation}
v²(R) = v^2(q) + 2 G m \biggl(\frac{1}{R} - \frac{1}{q}\biggr)
\label{vR}
\end{equation}
where $G$ is the constant of gravitation and $m$ is the mass of Saturn.

The collision velocity on Phoebe is computed assuming that the geometry of collisions 
is isotropic and then: 
\begin{equation}
v_i = \sqrt {v_p^2 + v_0^2}
\end{equation}
where $v_p$ is phoebe's orbital velocity and $v_0$ is the mean relative velocity of SDOs 
when they cross the orbit of Phoebe. This velocity was computed in the same way as 
$v(R)$ appropriately using Eq. (\ref{vR}) from $v(q)$ registered in our outputs.   
All the mentioned velocities and the radius and orbital velocity of Phoebe are shown in 
Table 1. 

\begin{table}
\begin{minipage}[t]{\columnwidth}
\caption { Radius $R_p$, density $\rho_p$, orbital velocity $v_p$ and 
superficial gravity times the mass  $g m_p$ of Phoebe. Mean relative velocity 
of SDOs when entering the Hill's sphere of Saturn $v(R)$, when they intersect the orbit of Phoebe 
 $v_0$ and  when collide with Phoebe  $v_i$.}
\label{Tabla1}
\centering
\renewcommand{\footnoterule}{}  
\begin{tabular}{l|c}
\hline 
 $R_p $ [km]             &   107.2      \\
 $\rho_p$ [gr cm$^{-3}$]   &  1.634  \\
$v_p$ [km s$^{-1}$]                 &  1.71     \\
$g m_p$ [km s$^{-2}$]                &  0.5532  \\
$v(R)$ [km s$^{-1}$]                 &  4.06    \\
$v_0 $[km s$^{-1}$]                   &  4.65    \\
$v_i$ [km s$^{-1}$]                 &  4.96    \\
\hline
\end{tabular}
\end{minipage}
\end{table}

Eq. (\ref{nc}) provides the number of collisions on Phoebe in relation to the 
number of encounters with Saturn that we had registered in our previous run. 

There are many papers based on theoretical and observational work which argue that the 
initial mass of the trans-Neptunian region was $\sim 100$ times greater than the present mass and decay 
to nearly its present value at most in 1 Gyr (see e.g. Morbidelli et al. \cite{Morby08}).
The simulation of Di Sisto \& Brunini (\cite{Disisto07}) studies the evolution of SDOs in the 
present configuration of the Solar System;
that is, when the SD is expected to have roughly reached its present mass and dynamic state, 
$\sim 3.5$ Gyrs ago. Then, we can estimate the  
total number of collisions on Phoebe during the last $\sim3.5$ Gyrs rescaling Eq. (\ref{nc}) in 
order to account for the total SDOs population.

Of the 1000 initial particles of our previous simulation (Di Sisto \&
Brunini \cite{Disisto07}), 368  underwent $10\, 257$ encounters within Saturn's Hill 
sphere. Therefore, the total number of encounters with Saturn of the whole SDO population 
in the present configuration of the Solar System  is estimated as
\begin{equation}
N_{et} = \biggl(\frac{368}{1000}\biggr) \;\; \biggl(\frac{10257}{368}\biggr) \,\, N, 
\end{equation}
where $N$ is the number of different SDOs which have existed, in the last 3.5 Gyrs and which can be 
obtained from Eq. (\ref{nr}). Here we assume that the present number of SDOs is roughly 
the same as it was 3.5 Gyrs ago.
Consequently, the total number of encounters with Saturn of the whole SDO 
population through the last 3.5 Gyrs depending on the diameter is given by:
\begin{equation}
N_{et} (d>d_0) = \biggl(\frac{10257}{1000}\biggr) \, N(d>d_0) 
\label{net}
\end{equation}
From this equation and Eq. (\ref{nc}) the total number of collisions of SDOs on Phoebe over 
the  last 3.5 Gyrs depending on the SDO's diameter, is given by:
\begin{equation}
N_{c}(d>d_0) =  \frac {v_i \, \,R_p^2} {v(R)\,\, R^2} \,\, N_{et} (d>d_0) 
\label{nct}
\end{equation}
Table \ref{tablaimp} shows some values of $N_c$ for certain diameters of the impactors.

Depending on the values of $s_2$, the diameter of the largest SDO impactor onto Phoebe 
during the last 3.5 Gyrs has been calculated as ranging from $110$ mts to $1.36$ km.

\begin{table}
\begin{minipage}[t]{\columnwidth}
\caption{Number of collisions of SDO impactors on Phoebe with diameters 
$d>d_0$ that produce craters with diameter $D>D_0$   ($N_c (D>D_0)$).
}
\label{tablaimp}
\centering
\renewcommand{\footnoterule}{}  
\begin{tabular}{|l|c|c|}
\hline 
$d_0 [km] $   & $D_0 [km] $  & $N_c (D>D_0)$      \\
\hline 
0.081          &  1        &  2 - 1180           \\
0.445          & 5            & 0 - 16           \\
0.969          & 10           & 0 - 2           \\
3.48          & 30           &  0             \\
\hline
\end{tabular}
\end{minipage}
\end{table}

\section{Craters on Phoebe by SDOs}
\label{sdo-crater}

The estimation of the size of a crater produced by a particular impact has been extensively studied. 
Schmidt \& Housen (\cite{SH87}) present a set of power-law scaling relations 
for the crater volume based on laboratory experiments that simulate crater formation and 
point-source solutions.  
Holsapple (\cite{H93}) also describe the scaling law for impact processes  in a review 
work.  The derived scaling laws allow us to link impacts of different sizes, velocities and 
superficial gravity and then obtain the size of a crater produced by a collision on a solar 
system body. Holsapple \& Housen (\cite{HH07}) present the updated scaling laws for cratering 
in a recent work dedicated to interpret the observations of the Deep Impact event. 
These cratering laws are used here to calculate the craters on Phoebe. Thus, the diameter $D$ of 
 a crater produced by an impactor of diameter $d$ can be obtained from the general equation 
(Holsapple \& Housen \cite{HH07}):  
\begin{equation}
D = K_1 \left[ \left(\frac{g d}{2 v_i^2}\right) \left(\frac{\rho_t}{\rho_i}\right)
^{\frac{2 \nu}{ \mu}} +  \left(\frac{ Y}{\rho_t v_i^2}\right)^{\frac{2+\mu}{2}} \left(\frac{\rho_t}{\rho_i}\right)^{\frac{\nu (2+\mu)}{\mu}} \right ]^{-\frac{\mu}{2+\mu}}   d
\label{ds}
\end{equation}  
$\rho_t$ being the target density, $g$ its superficial gravity, Y its strength, 
$\rho_i$  the density of the impactor and  $v_i$  the collision velocity.  
This  impact cratering scaling law depends  on two exponents, $\mu$ and $\nu$, and a constant, $K_1$,
that characterize the different materials.   
The first term in the square brackets is a measure 
of the importance of gravity in the cratering event and the second is a measure 
of the importance of the target strength. Thus, if the first term dominates on the second 
term, the crater is under the gravity regime, and if the second term dominates we have the 
strength regime. The partition between the two size scales of impacts depends on the size of 
the event  (Holsapple \cite{H93}).  
Eq. (\ref{ds}) is a convenient empirical smoothing function to span the transition 
between the gravity regime and the strength regime (Holsapple, \cite{H93}). 
Since Phoebe is a small satellite with a relatively low gravity, 
the strength regime can be important for the smaller craters. 
 
As Phoebe's density ($1.63 gr/cm^3)$ is similar to sand and lower compacting 
material is found on its surface, we adopt $K_1= 1.03$, $\mu = 0.41 $  and $\nu=0.4$ that 
correspond to sand or cohesive soil in Holsapple \& Housen (\cite{HH07}).
This value of $\mu$ corresponds to materials  with a porosity of 
$\sim 30 -35 \%$ (Holsapple \& Schmidt  \cite{HS87}) which is compatible with the ranges of 
Phoebe's predicted porosity. The value for dry soils from Holsapple (\cite{H93}), i.e. $Y=0.18$ mpa,   
is used for the strength.

The calculated densities of TNOs vary considerably from  $\sim 0.5 - \sim 3$ $gr/cm³$. 
 Although  a dimension-density trend  has been suggested (Sheppard et al.  
\cite{S08}, Perna et al.  \cite{P09}), more data are required to confirm it.
 In addition; as crater experiments do not include variations in the impactor 
material, there is no data to precisely determine the  dependence on the impactor density 
(Schmidt \& Housen \cite{SH87}, Housen \& Holsapple \cite{HH03}). 
Therefore, we assume $\rho_i = \rho_t$, which is also between the ranges of 
calculated densities in the trans-Neptunian region. 
By taking all this into account,  the diameter of a crater on Phoebe  for a given impactor 
diameter can be calculated  from Eq. (\ref{ds}) through:
\begin{equation}
D = 1.03 \left[ \left(\frac{g d}{2 v_i^2}\right) +  \left(\frac{ Y}{\rho_t v_i^2}\right)^{1.205} 
 \right ]^{-0.17}   d
\label{dsp}
\end{equation}  
This equation describes simple bowl-shape craters but, as mentioned in the introduction, 
Cassini images of Phoebe reveal basically simple crater shapes with the only 
exception of Jason with a diameter of $\sim 100$ km (Giese et al. \cite{Giese06}). 
 Hence we use Eq. (\ref{dsp}) for calculating the diameters of all craters on Phoebe 
  without any further correction for transient-to-final size.
By combining  Eq. (\ref{nct}) and (\ref{dsp}) it is possible to calculate 
the number of craters on Phoebe according to the diameter of the crater. 
 Fig. (\ref{fig1})  shows the cumulative number of craters, with diameters greater 
than a given value for the two size distribution power laws for smaller SDOs on Phoebe. Note that the 
different slopes in the number of craters for each curve is due to the difference in both 
 indexes $s_2$ considered. 
As we mentioned before, there is a limit impactor diameter that accounts for the 
transition between the gravity regime and the strength regime. This diameter can be  
 obtained equating the first and second terms of Eq. (\ref{dsp}). This limit impactor diameter 
is $d_l=367 mts$, which produce a limit crater of 
$D_l=4.2$ km. Thus, for crater diameters $D < D_l$, the production of craters on Phoebe is under 
the strength regime and for  $D > D_l$, the production of craters  is under 
the gravity regime. 
 Note that  $D_l$ depends strongly on the assumed value of strength which is actually unknown. 
Then considering other values of the strength for less cohesive soils as terrestrial dry desert 
alluvium of $Y = 65 $ kpa (Holsapple \& Housen \cite{HH07}) and Surface Lunar regolith $Y = 10$ kpa 
(Holsapple \cite{H11}), $D_l$ could take the values 1.5 km and 233 mts respectively. 
We will consider in the following that $D_l = 4.2$ km but it must be taken into account that 
$D_l$ can be as small as 233 mts.

Since in the strength regime the crater diameter depends  linearly on 
the impactor diameter, the relation between the cumulative number of craters on Phoebe 
and the crater diameter follows the same power law relation as that followed by the number of SDOs. 
For $D < 4.2$ km the cumulative number of craters on Phoebe follows a power law with a 
cumulative index of 1.5 and 2.5, according to the value of $s_2 = 2.5$ or $s_2=3.5$ respectively.   
For $D>4.2$ km, this is in the gravity regime, the crater diameter does not depend linearly on 
the impactor diameter. Therefore, we  fit a power law for the cumulative number of craters 
on Phoebe depending on the crater diameter of index $2.8$. 
Kirchoff \& Schenk (\cite{KS10}) found  that the crater size frequency distribution for Phoebe has 
a cumulative index of $2.348$ for $D = 0.15 - 1$ km and $1$ for $D = 1-4$ km. We can see that, 
for very small craters, this index is very similar to our value of $s_2 = 3.5$ or cumulative index 
of $2.5$. This is consistent with the fact that  the size distribution of very small   
objects is expected to approach a Donhanyi size distribution ($s_2=3.5$) and then the 
craters produced by those small projectiles (that are in the strength regime) have to follow 
the same power law size distribution.
Besides Kirchoff \& Schenk (\cite{KS10}) found that for $D = 1-4$ km, Phoebe's crater distribution 
has a shallow slope and this  implies that Phoebe has a deficiency of craters with $D\sim 1.5$ 
km. This change of slope cannot be explained by our method and our proposed contribution 
of Centaurs from the SD to the craters on Phoebe,  unless that the considered SDF of SDOs 
is different. Anyway  more work in relation to another 
source of craters on Phoebe, as planetocentric objects, is needed and also  
its connection to the origin of the irregular satellite itself.

According to the differential size distribution 
index $s_2$, the largest crater on Phoebe produced by a Centaur from the SD has
a diameter of between $1.4 $ km and $13.5$ km.  
Table \ref{tablaimp} shows the cumulative number of craters 
on Phoebe greater than certain diameters produced by Centaurs from the SD in the current
configuration of the Solar System  during the last 3.5 Gyrs. Since the largest crater on Phoebe 
has a diameter of $\sim 100$ km, it is unlikely that it was produced by a recent collision of 
an SDO. This will be discussed in a following section.

\begin{figure}
\centering
\resizebox{\hsize}{!}{\includegraphics{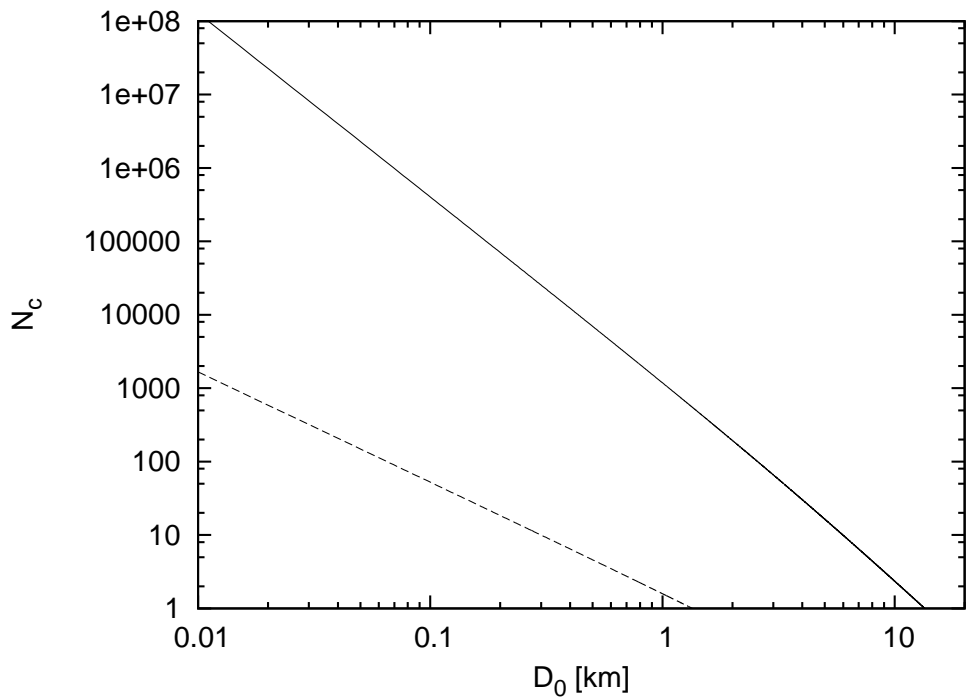}}
\caption{ Cumulative number of craters  with diameters greater than $D_{0}$ in the last 3.5 Gyrs,  
produced by  Centaurs from SDOs on Phoebe. Filled line corresponds to the  differential 
power law index $s_2 = 3.5$ and the dotted line to $s_2 = 2.5$ }
\label{fig1}
\end{figure}

\section{Rate of SDO collisions on Phoebe}

\begin{figure}
\centering
\resizebox{\hsize}{!}{\includegraphics{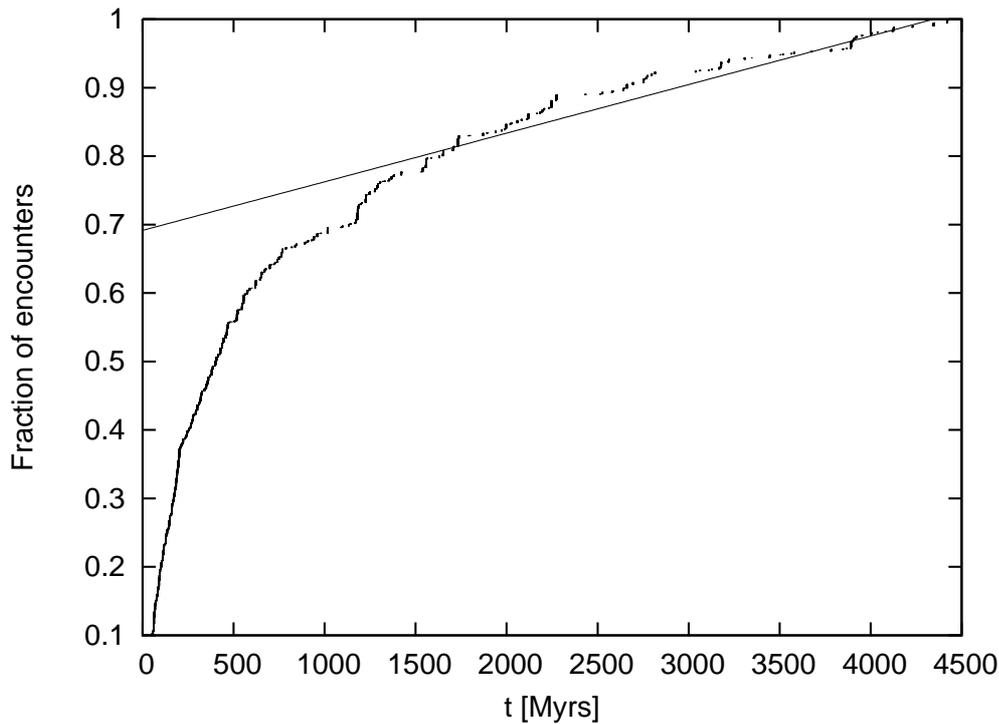}}
\caption{ Fraction of encounters of SDOs with Saturn. 
The linear fit to the data the last 3.5 Gyrs is also shown.}
\label{tasae}
\end{figure}

From our outputs we can calculate the number of encounters within the Hill's sphere of 
Saturn as a function of time. In Fig. \ref{tasae}  is plotted the normalize cumulative 
number of encounters as a function of time.  The whole plot can be fitted by a log-function 
given by $f(t) =  a\,\, + b \,\, log\, t$, where $a = -3.24$ and $b=0.19$. 
The total cumulative number of encounters with Saturn for all diameters for each time
can be obtained from the plot and Eq. (\ref{net}). 
We have calculated the number of collisions on Phoebe from the number of encounters with 
Saturn from our outputs  (Eq. \ref{nct}), and this number can be calculated for each time 
in the integration. 
Then the cumulative number of collisions (or cumulative number of craters) on Phoebe for all 
diameters as a function of time can be obtained  by multiplying the fraction of encounters 
obtained from Fig. \ref{tasae} by the number of collisions $ N_{c} (d>d_0)$.

As we can see  from Fig. \ref{tasae} the rate of encounters and then the rate of collisions on 
Phoebe was high  at the beginning but it has been  decreasing up to the present. 
 The first Myrs  the shape of the curve  is purely arbitrary due to initial conditions
but then it will be significant and is stabilizing.
In the last $\sim 3.5 $ Gyrs the rate has been almost constant. 
In fact,  it is possible to fit a linear relation to the last 
3.5 Gyrs of Fig. \ref{tasae}  given by $g(t)= \dot F t+ c $, where $\dot F = 7.1\times 10^{-11}$ 
and $c = 0.69$.  This linear approximation allows us to calculate a present rate of craters for a 
given diameter. The slope of this linear function $\dot F = 7.1 \times 10^{-11}$ is
the present normalized rate of encounters of SDOs with Saturn  per year. 
To obtain the present rate of encounters with Saturn for each diameter we must multiply $\dot F$ by 
$N_{et} (d>d_0)$ (obtained from Eq (\ref{net})). And then the present rate of collisions on Phoebe
for each diameter can be obtained  multiplying $\dot F$ by 
 $N_{c} (d>d_0)$ (obtained from Eq (\ref{nct})). 
Similarly, the current rate of cratering on Phoebe for craters greater than a given diameter 
can be obtained from the current rate of collisions and the relation (\ref{dsp}) between  
the diameter of the impactor and the diameter of the crater. 
Thus, for example the current cratering rate on Phoebe of Centaurs from SDOs that produces craters 
with $D > 1$ km is between  $1.4 \times 10^{-10}$ and $8.3 \times 10^{-8}$ craters per year 
(depending on the $s_2$ value); this is at least $\sim 80$ craters with $D > 1 $ km in  the 
last Gy.  The current cratering rate of craters with $D > 5$ km is at least $1.14 \times 10^{-9}$ 
craters per year, this is  $\sim 1$ crater with $D > 5 $ km in the last  Gy. 

Zhanle et al. (\cite{Z03}) calculate cratering rates in the satellites of the outer planets. 
They obtained a cratering rate on Phoebe for craters with $D > 10$ km of $8.6 \times 10^{-11} 
year^{-1}$. In this study we obtain a cratering rate on Phoebe of craters with $D > 10$ km 
of between $2.7 \times 10^{-12}$ and $1.4 \times 10^{-10}$ 
craters per year, depending on $s_2$. Our value for $s_2 = 3.5$ is  very similar to the  
previous calculation of Zhanle et al. (\cite{Z03}).

\section{The contribution of escaped Plutinos to the craters on Phoebe}

Plutinos could be another source of craters on Phoebe. In a recent paper, Di Sisto et al. (\cite{Disisto10}) 
study the post escape evolution of Plutinos when they escaped from the  3:2 mean 
motion resonance with Neptune, and in particular 
their contribution to the population of Centaurs. In that work, we perform two sets  of 
numerical simulations in order first to identify the plutinos that have recently escaped from the 
resonance and second to follow their evolution under the influence of the Sun and the four giant 
planets. This numerical simulation considers the evolution of plutinos in the present configuration 
of the 
Solar System, this is as we did for SDOs, when the trans-Neptunian region is expected to have reached 
roughly its present mass and dynamical state, $\sim 3.5$ Gyrs ago. Following the same analysis that 
we made for SDOs described in Sects. \ref{sdo-col} and \ref{sdo-crater}, we calculate the number 
of craters on Phoebe from escaped plutinos the last 3.5 Gyrs, and also the greater impactor and crater. 
In the numerical simulations by Di Sisto et al. (\cite{Disisto10}), we start with $20\,000$  initial 
massless particles in the  3:2 mean motion resonance, 671 of them undergone $20\, 459$ encounters 
within Saturn's Hill sphere during the integration. We obtained that the mean relative encounter 
velocity of plutinos when entering the Hill's sphere of Saturn is  $v(R)= 4.57$ $km/s$, the mean 
relative velocity of plutinos when they intersect the orbit of Phoebe is $v_0 = 5.12 $ $km/s$ and the 
collision velocity of 
plutinos on Phoebe is  $v_i = 5.4 $ $km/s$.  We take the present  number of plutinos from 
 de El\'{\i}a et al. (\cite{deelia08}), but consider that the size distribution breaks at $d \sim 60$ km
with the two limit power law indexes $ s_2 = 3.5$ and $2.5$ as we adopted for SDOs to be consistent 
(see Sect. \ref{ssdo}). 
Those indexes give the highest and lowest number of SDOs and then the highest and lowest 
production of craters on Phoebe. 
Then the present cumulative number of plutinos is given by:
\begin{xalignat}{4}
N(>D) &= C \bigg(\frac{1 \text{km}}{D}\bigg)^{p} &&\text{for} && D \leq 60 \text{km},  \nonumber \\
N(>D) &= 7.9 \times 10^{9} \bigg(\frac{1 \text{km}}{D}\bigg)^{3} &&\text{for} && D > 60 \text{km},
\label{e33}
\end{xalignat} 
where $C$ = 7.9 $\times$ 10$^{9}$ (60)$^{p-3}$ by continuity for $D$ = 60 km and the  
cumulative power law index $p$ adopt the values $2.5$ and $1.5$ ($p=s-1$). This law is plotted 
in Fig 1.

Considering all this and our method described above, we obtained that the greater escaped plutino 
impactor on Phoebe has a diameter between  $1.5$ mts and $102$ mts that produces a crater  
between $19.3$ mts 
and $1.3$ km respectively, depending on the power index $p$ of the size distribution of plutinos.
Also, we can obtain the  number of craters on Phoebe from escaped plutinos. We have at least 
two craters greater than $1$ km on Phoebe from plutinos.   
Comparing this with the values obtained for the contribution of SDOs, it can be stated that the 
number of craters 
produced by escaped plutinos on Phoebe is negligible with respect to the contribution of SDOs. 
Also  it can be stated that the greater craters are signed by the contribution of SDOs.

\section{Discussion}

In the previous sections, we have calculated the production of craters on Phoebe considering the 
present population in the SD and Plutinos. However - as mentioned - there are many papers based 
on theoretical and observational work that argue that the initial mass of the TN region was 
$\sim 100$ times 
greater than the present mass (see e.g. Morbidelli et al. \cite{Morby08}). Observations predict 
a current mass of the Kuiper Belt that is very small with respect to that required for models to 
grow the objects that we see. The mass depletion due to a strong dynamical excitation of the 
Kuiper Belt is thought to be the scenario for this ``mass deficit problem''.  There were several 
models that try to describe the mass depletion; the last model 
that described this mechanism is the ``Nice model'' where the Kuiper Belt had to be significantly 
depleted before the time of the LHB (Levison et al. \cite{niza08}).
The ``Nice Model'' assumes the giant planets initially in a more compacted region from $\sim 5.5 $ to 
$\sim 14$  AU and a planetesimal disk of a total mass of $\sim 35$ $M_T$ that extends beyond the orbits 
of the giant planets up to $\sim 34$ AU. The interaction between the planets and planetesimals makes 
the giant planets migrate for a long time removing particles from the system. After a time ranging from 
350 My to 1.1 Gy, Jupiter and Saturn cross their mutual 1:2 mean motion resonance. 
Then, the eccentricities of Uranus and Neptune drives up and those planets penetrate into the 
planetesimal disk. 
This destabilizes the  full disk and the planetesimals are scattered all over the Solar System.  

Beyond the model and the mechanism responsible for the mass depletion of the trans-Neptunian zone, 
 we can consider that  primitive SDOs  (that were 100 times  more numerous than the present ones) 
follow the same dynamical evolution than the present population when they enter the 
planetary zone as Centaurs. Then we can 
 calculate in the same way as we did in the previous sections  and with the same model 
the cratering on Phoebe assuming an initial population of SDOs 100 times the present population. 
 
This is an estimation since we have to know the real initial scenario of formation 
of the Solar System and in particular of SDOs. However when a SDO enter the Centaur zone, 
inside the orbit of Neptune, its dynamical evolution is governed by the giant planets and then 
the particular initial scenario can be considered secondary for the present study. 

Doing that, we obtain that the greater impactor on Phoebe during the age of the Solar System has a 
diameter between 2.4 to 8.6  km and produces a crater of 21.6 to 64.2 km. 
The value  corresponding to $s_2=3.5$   (64.2 km) is in agreement (within the expected 
errors and statistical fluctuations) with the observation of 
the great crater ``Jason'' on Phoebe with a diameter of $ \sim 100$ km. 
The number of craters greater than a given diameter can be obtained increasing 100 times the 
values obtained in Sect. \ref{sdo-crater}

Recently, Cassini images of Phoebe have allowed to study its surface and the craters in it. 
Kirchoff \& Schenk (\cite{KS10}) obtain,  
 a cumulative crater density for $D \ge 5$ km of $2233 \pm 1117$ based on crater counting
 from  Cassini images. 
From  our model, assuming an initial number of SDOs 100 times the present population, 
 we obtain  $ N_{c}(D > 5 km) = 12 - 1640$, again  
in a good agreement with the values obtained by Kirchoff \& Schenk (\cite{KS10}) 
 $s_2 = 3.5$.

\section{Conclusion}

We have studied the production of craters on Phoebe from SDOs and escaped plutinos that have 
reach the Saturn zone in the present configuration of the Solar System. 
We have obtained that the contribution of escaped plutinos is negligible with respect to the 
contribution of SDOs. We have obtained that both the number of craters and the greater crater on
 Phoebe
produced by SDOs cannot account to the observations. But if we take into account that 
the initial mass of the trans-Neptunian region was 100 times the present one, we match the 
craters produced by SDOs on Phoebe with the observed characteristics of the satellite 
if $s_2 = 3.5$. 

Those considerations imply that the main cratering features of Phoebe must be acquired  
when the SD had being depleted at the early times of evolution of the Solar System. 
More than that, if the ``Nice model'' describes correctly the scenario of the origin of 
the Solar System, the scattering inward of planetesimals by Neptune and Uranus in that model 
must  be similar to the present scattering in our model, and the TNOs have to lose 
 memory when they arrive to Saturn.

If this is what happened and the main crater characteristics on Phoebe were produced when Phoebe 
was a satellite of Saturn, 
the great agreement of our model with the observations constrain the 
time when Phoebe had to be captured, very early in the  evolution of our Solar System. This 
was also suggested by Levison et al. (\cite{niza08}). 

We have obtained that the present normalized rate of encounters of SDOs 
with Saturn is: $\dot F = 7.1 \times 10^{-11}$  per year. From this number we could obtain 
 the present cratering rate on Phoebe for each crater diameter.

We have compared the size crater distribution on Phoebe obtained from our model  
 with the observations of craters by Kirchoff \& Schenk (\cite{KS10}). 
Our crater size frequency distribution agree with that 
obtained by Kirchoff \& Schenk for very small impactors that produce craters with 
 $D = 0.15 - 1$ km. This distribution follow a power law with a 
cumulative index of $2.5$ consistent with a Donhanyi size distribution. 
For craters of $D = 1-4$ km  Kirchoff \& Schenk (\cite{KS10}) found 
a shallow slope and  a deficiency of craters with $D\sim 1.5$ km. 
This change of slope cannot be explained by our method and the contribution 
of Centaurs from the SD. More work in relation to another 
source of craters on Phoebe, as planetocentric objects is needed, and also  
its connection to the origin of the irregular satellite itself.

\vspace*{0.5cm}

\noindent {\bf Acknowledgments:} We thank  Gonzalo de El\'{\i}a for valuable discussion on this paper. 
We also acknowledge  an anonymous referee that made valuable comments and suggestions 
that helped us to improve the manuscript.

\end{document}